\documentclass[fleqn,usenatbib]{mnras}

\usepackage{newtxtext,newtxmath}

\usepackage[T1]{fontenc}

\DeclareRobustCommand{\VAN}[3]{#2}
\let\VANthebibliography\thebibliography
\def\thebibliography{\DeclareRobustCommand{\VAN}[3]{##3}\VANthebibliography}

\usepackage{graphicx}	
\usepackage{amsmath}	

\title[GPU-accelerated nested sampling for gravitational waves]{Gravitational-wave inference at GPU speed: A \texttt{bilby}-like nested sampling kernel within \texttt{blackjax-ns}}

\author[Metha Prathaban et al.]{
Metha Prathaban,$^{2,3}$\thanks{E-mail: myp23@cam.ac.uk (MP)}
David Yallup,$^{1,2}$
James Alvey,$^{1,2}$
Ming Yang,$^{1}$
Will Templeton,$^{1}$
and Will Handley$^{1,2}$
\\
$^{1}$Institute of Astronomy, University of Cambridge, Cambridge, CB3 0HA, UK\\
$^{2}$Kavli Institute for Cosmology, University of Cambridge, Cambridge, CB3 0EZ, UK\\
$^{3}$Department of Physics, University of Cambridge, Cambridge, CB3 0HE, UK\\
}

\date{Accepted XXX. Received YYY; in original form ZZZ}

\pubyear{\the\year{}}

\begin{document}
\label{firstpage}
\pagerange{\pageref{firstpage}--\pageref{lastpage}}
\maketitle

\begin{abstract}
We present a GPU-accelerated implementation of the gravitational-wave
Bayesian inference pipeline for parameter estimation and model comparison. 
Specifically, we implement the `acceptance-walk'
sampling method, a cornerstone algorithm for gravitational-wave
inference within the \texttt{bilby} and \texttt{dynesty} framework.
By integrating this trusted kernel with the vectorized \texttt{blackjax-ns}
framework, we achieve typical speedups of 20-40x for aligned spin binary black hole
analyses, while recovering posteriors and evidences that are
statistically identical to the original CPU implementation. This faithful
re-implementation of a community-standard algorithm establishes a foundational
benchmark for gravitational-wave inference. It quantifies the
performance gains attributable solely to the architectural shift to GPUs,
creating a vital reference against which future parallel 
sampling algorithms can be rigorously assessed. 
This allows for a clear distinction between algorithmic innovation 
and the inherent speedup from hardware.
Our work provides a validated community tool for performing
GPU-accelerated nested sampling in gravitational-wave data analyses.
\vspace{0.5cm}
\end{abstract}




\section{Introduction}

The era of gravitational-wave (GW) astronomy, initiated by the landmark
detections at the Laser Interferometer Gravitational-Wave Observatory (LIGO),
Virgo and now KAGRA, has revolutionized
our view of the cosmos~\citep{GW150914, GW170817,GWTC1,GWTC2, GWTC3,GWTC3_pop_analysis,GWTC2_GR,siren}. 
Extracting scientific insights from the data, from measuring the
masses and spins of binary black holes to performing precision tests of general
relativity, relies heavily on the framework of Bayesian inference~\citep{Thrane_2019}. 
This allows for the estimation of posteriors on source parameters
(parameter estimation) and the comparison of competing physical models (model selection).

The process of Bayesian inference in GW astronomy is, however, computationally
demanding. Realistic theoretical models for the GW waveform can be complex, and the
stochastic sampling algorithms required to explore the high-dimensional parameter
space can require millions of likelihood evaluations per analysis~\citep{LIGO_guide_signalextraction}.
There are several community-standard software tools for performing GW inference, such as
\texttt{pycbc}~\citep{pycbc} and \texttt{lal}~\citep{lal}.
Here, we focus on the inference library \texttt{bilby}~\citep{bilby_paper}
paired with a custom implementation of the nested sampler \texttt{dynesty}~\citep{dynesty}, 
which has proven to be a robust and highly effective framework, tuned for the specific needs of GW posteriors.
However, this framework is predominantly executed on central processing units
(CPUs), making individual analyses costly and creating a
computational bottleneck. This challenge is set to become acute with the
increased data volumes from future observing runs~\citep{aLIGO, aVirgo, aLVK_prospects} and the advent of
next-generation observatories, such as the Einstein Telescope~\citep{ET_science_case}, which promise
unprecedented sensitivity and detection volumes~\citep{HuAccelerationReview}.

In response to this challenge, the GW community has begun to leverage the immense
parallel processing power of graphics processing units (GPUs). Pioneering work in
this domain, such as the \texttt{jimgw} codebase~\citep{wong2023fastgravitationalwaveparameter}, has successfully
implemented the GPU-accelerated Markov Chain Monte Carlo (MCMC) sampler \texttt{flowMC}~\citep{flowMC}, 
paired with GPU implementations of waveform models provided by the
\texttt{ripple} library~\citep{ripple}. This work has demonstrated that
substantial, orders-of-magnitude speedups in runtime are
achievable for GW parameter estimation. This success in accelerating
parameter estimation motivates the development of complementary
GPU-native algorithms to accelerate other key inference tasks, such as 
model selection.

In this paper, we introduce a GPU-accelerated nested sampling (NS)
algorithm for gravitational-wave data analysis. Our method
builds upon the trusted `acceptance-walk' sampling method used in the
community-standard \texttt{bilby} and \texttt{dynesty} framework~\citep{bilby_paper}.
We leverage the \texttt{blackjax}~\citep{jax2018github,cabezas2024blackjax} implementation of nested sampling~\citep{yallup2025nested}, 
a reformulation of the core nested sampling algorithm for massive GPU parallelization. 
In place of the default method for particle evolution recommended in~\cite{yallup2025nested}, we develop a custom
kernel that implements the `acceptance-walk' method.
This ensures our sampler's logic is almost identical to the
\texttt{bilby} and \texttt{dynesty} implementation, with differences
primarily related to parallelization. This approach offers
\texttt{bilby} users a direct path to GPU-accelerated inference for the most expensive problems in GW astronomy.
They can achieve significant speedups while retaining the same
robust and trusted algorithm at the core of their analyses.

A key motivation for this work is to establish a clear performance
baseline for GPU-based nested sampling in gravitational-wave astronomy.
By adapting the community-standard \texttt{bilby}
`acceptance-walk' sampler for a GPU-native framework, we aim to isolate
and quantify the speedup achieved from hardware parallelization alone.
This provides a crucial reference point, enabling future work on novel
sampling algorithms to be benchmarked in a way that disentangles
algorithmic improvements from architectural performance gains.

In the following section, we summarise the core ideas
of Bayesian inference, nested sampling and GPU architectures.
We then describe the implementation of the `acceptance-walk' sampling
method in the \texttt{blackjax-ns} framework in Section~\ref{sec:methods}, and validate it against
the \texttt{bilby} and \texttt{dynesty} implementation in Section~\ref{sec:results}, discussing
our results. Finally, we present our conclusions in Section~\ref{sec:conclusions}.

\begin{figure*}
    \centering
    \includegraphics{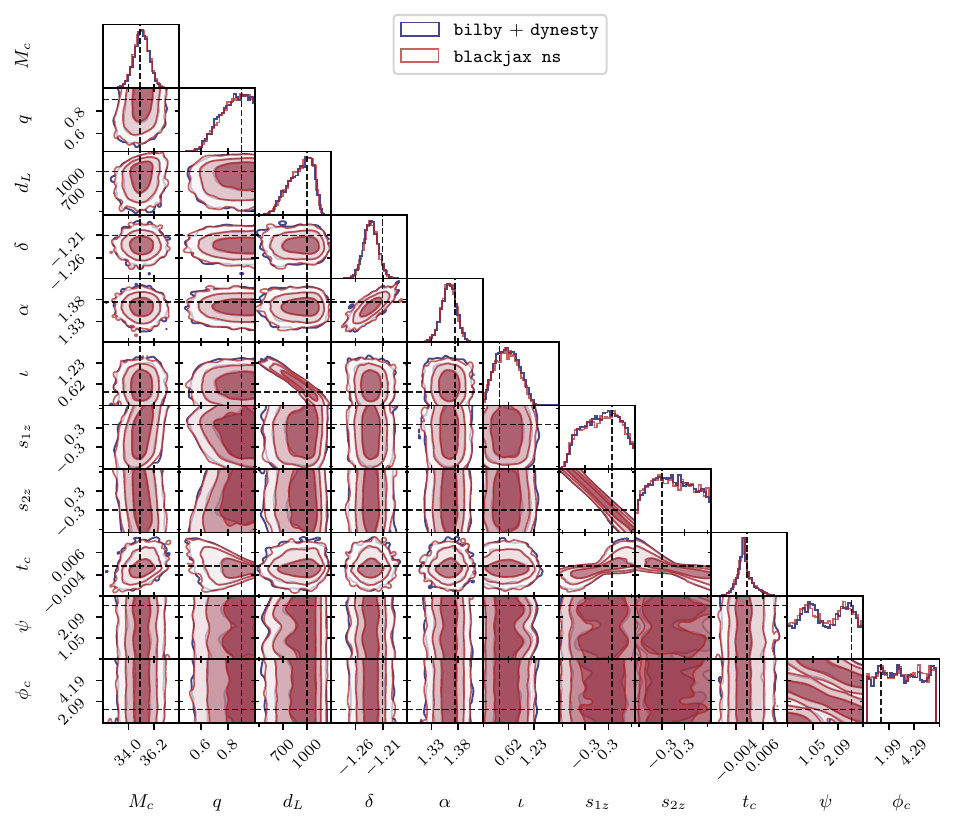}
    \caption{Recovered posteriors for the 4s signal. The posteriors are in excellent agreement with each other, demonstrating that the
    \texttt{blackjax-ns} implementation with our custom kernel is functionally equivalent to the \texttt{bilby} + \texttt{dynesty} implementation.}
    \label{fig:4s_posteriors}
\end{figure*}

\section{Background}
\label{sec:background}

\subsection{Bayesian inference in GW astronomy}
\label{sec:background_bayes}
We provide a brief overview of the core concepts of
Bayesian inference as applied to GW astronomy. For a more
comprehensive treatment, we refer the reader to~\cite{skilling, Thrane_2019, lal, bilby_paper}
and~\cite{LIGO_guide_signalextraction}.

The analysis of GW signals is fundamentally a problem of statistical
inference, for which the Bayesian framework is the community standard.
The relationship between data, $d$, and a set of parameters,
$\theta$, under a specific hypothesis, $H$, is given by Bayes' theorem~\citep{Bayes1763}:

\begin{equation}
    \mathcal{P}(\theta|d, H) = \frac{\mathcal{L}(d|\theta, H) \pi(\theta|H)}{Z(d|H)}.
    \label{eq:bayes_theorem}
\end{equation}%
Here, the posterior, $\mathcal{P}(\theta|d, H)$, is the probability distribution
of the source parameters conditioned on the observed data. It is
determined by the likelihood, $\mathcal{L}(d|\theta, H)$, which is the
probability of observing the data given a specific realisation of the
model, and the prior, $\pi(\theta|H)$, which encodes initial beliefs
about the parameter distributions.

The denominator is the Bayesian evidence,
\begin{equation}
    Z(d|H) = \int \mathcal{L}(d|\theta, H) \pi(\theta|H) d\theta,
    \label{eq:evidence}
\end{equation}
defined as the likelihood integrated over the entire volume of the
prior parameter space.

There are two pillars of Bayesian inference of particular interest
in GW astronomy. The first, parameter estimation, seeks to infer 
the posterior distribution $\mathcal{P}(\theta|d, H)$ of the source parameters of a signal or population of signals.
The second, model selection, evaluates two competing
models, $H_1$ and $H_2$, under a fully Bayesian framework by computing the ratio of
their evidences, known as the Bayes factor, $Z_1 / Z_2$. This enables
principled classification of noise versus true signals, as well as the 
comparison of different waveform models~\citep{LIGO_guide_signalextraction}.

In GW astronomy, the high dimensionality of the parameter space and
the computational cost of the likelihood render the direct evaluation
of Eq.~\ref{eq:bayes_theorem} and Eq.~\ref{eq:evidence} intractable~\citep{LIGO_guide_signalextraction}.
Analysis therefore relies on stochastic sampling algorithms to
numerically approximate the posterior and evidence.

\subsection{GPU-accelerated nested sampling}

\label{sec:background_ns_and_gpus}

\subsubsection{The nested sampling algorithm}
\label{sec:background_ns}

Nested Sampling is a Monte Carlo algorithm designed to solve the
Bayesian inference problem outlined in Sec.~\ref{sec:background_bayes}.
A key strength of the NS algorithm is that it directly computes the
Bayesian evidence, $Z$, while also producing posterior samples as a
natural by-product of its execution~\citep{skilling, dynamic_ns}.

The algorithm starts by drawing a population of $N$ `live points'
from the prior distribution, $\pi(\theta|H)$. It then proceeds
to iteratively evolve this population. In each iteration, the live point with the lowest
likelihood value, $\mathcal{L}_{\text{min}}$, is identified. This point
is deleted from the live set and stored. It is then replaced with a
new point, drawn from the prior, but subject to the hard constraint that
its likelihood must exceed that of the deleted point,
i.e., $\mathcal{L}_{\text{new}} > \mathcal{L}_{\text{min}}$. This process
systematically traverses nested shells of increasing likelihood, with
the sequence of discarded points mapping the likelihood landscape.

The primary computational challenge within the NS algorithm is the
efficient generation of a new point from the likelihood-constrained
prior~\citep{NSNature}. The specific method used for this so-called `inner kernel' is a
critical determinant of the sampler's overall efficiency and robustness~\citep{NS_methods_buchner}.

\subsubsection{GPU architectures for scientific computing}
\label{sec:background_gpus}

The distinct architectures of Central Processing Units (CPUs) and
Graphics Processing Units (GPUs) offer different advantages for
computational tasks. CPUs are comprised of a few powerful cores
optimised for sequential task execution and low latency. In contrast,
GPUs feature a massively parallel architecture, containing thousands of
simpler cores designed for high-throughput computation~\citep{GPU_computing}.

This architecture makes GPUs exceptionally effective for problems that can
be expressed in a Single Instruction, Multiple Data (SIMD) paradigm~\citep{CUDAGuide}.
In such problems, the same operation is performed simultaneously across
a large number of data elements, leading to substantial performance
gains over serial execution on a CPU. The primary trade-off is that
algorithms must be explicitly reformulated to expose this parallelism,
and not all computational problems are amenable to vectorization.
There is algorithmic overlap with vectorisation and multi-threading on CPUs~\citep{Handley:2015vkr, Smith:2019ucc}; 
nonetheless, the GPU execution model is uniquely suited to truly massive parallelism. We seek to exploit this 
capability in the present work.

\subsubsection{A vectorized formulation of nested sampling}
\label{sec:background_vectorized_ns}

The iterative, one-at-a-time nature of the traditional NS algorithm
is intrinsically serial, making it a poor fit for the parallel
architecture of GPUs. To overcome this limitation,~\cite{yallup2025nested}
recently developed a vectorized formulation of the NS algorithm,
specifically designed for highly parallel execution within the
\texttt{blackjax} framework~\citep{cabezas2024blackjax}.

One of the core parts of this approach is the idea of batch
processing. Instead of replacing a single live point in each iteration,
the algorithm removes a batch of $k$ points with the lowest likelihoods
simultaneously~\citep{parallel_ns}. The critical step of replacing these points is then
parallelized. The algorithm launches $k$ independent sampling processes
on the GPU, with each process tasked with finding one new point that
satisfies the likelihood constraint, $\mathcal{L} > \mathcal{L}_{\text{min}}$,
where $\mathcal{L}_{\text{min}}$ is now the maximum likelihood of the
discarded batch.

This reformulation transforms the computationally intensive task of
sample generation from a serial challenge into a massively parallel one,
thereby leveraging the architectural strengths of the GPU. 
Design and implementation of massively parallel Markov Chain Monte Carlo 
(MCMC) kernels is a non-trivial task, and an area of active 
research~\citep{pmlr-v151-hoffman22a,pmlr-v130-hoffman21a}. 
The original work of \cite{yallup2025nested} proposed a 
Hit-and-Run Slice Sampling~\citep{Neal2003_slice} kernel for this task, 
alongside providing a general vectorized framework which we leverage in this work. 
Here, we implement one particular inner kernel used within the \texttt{bilby}
and \texttt{dynesty} framework.
Establishing this baseline demonstrates the flexibility of the \texttt{blackjax} 
nested sampling framework, and gives scope for future work to explore other 
parallel inner sampling methods.

\vspace*{-20pt}
\section{Methods}
\label{sec:methods}

\subsection{The inner sampling kernel}
\label{sec:methods_kernel}

\begin{figure}
    \centering
    \includegraphics{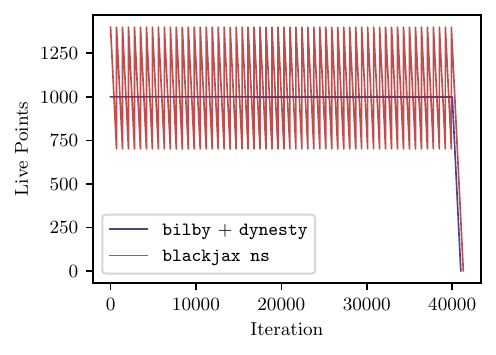}
    \caption{Comparison of the number of live points in the sequential CPU and batched GPU implementations.
    Although the nominal number of live points used in \texttt{blackjax-ns} is higher than in \texttt{bilby},
    the saw-tooth pattern means that the effective number of live points is the same. Here, we use a batch size of $k = 0.5 \times n_{\text{live}}$.}
    \label{fig:nlive_comparison}
\end{figure}

Several inner sampling methods are implemented within the
\texttt{bilby} and \texttt{dynesty} framework~\citep{bilby_paper, dynesty}. In this work, we
focus on a GPU-accelerated implementation of the `acceptance-walk'
method, which is a robust and widely used choice for GW analyses.

In the standard CPU-based \texttt{dynesty} implementation, the sampler
generates a new live point by initiating a Markov Chain Monte Carlo
(MCMC) walk from the position of the deleted live point. 
The proposal mechanism for the MCMC walk is based on Differential
Evolution (DE)~\citep{DE, DE2}, which uses the distribution of existing live points to
inform jump proposals. A new candidate point is generated by adding a
scaled vector difference of two other randomly chosen live points to the
current point in the chain. Under the default \texttt{bilby} configuration, 
the scaling factor for this vector is chosen stochastically: with equal probability, 
it is either fixed at 1 or drawn from a scaled gamma distribution\footnote{
$\text{Drawn from} \frac{2.38}{\sqrt{2 n_{\textrm{dim}}}} \times \Gamma(4,0.25)$,
where $n_{\textrm{dim}}$ is the parameter space dimensionality.}. This proposed 
point is accepted if it satisfies the likelihood constraint, $\mathcal{L} > \mathcal{L}_{\text{min}}$,
where $\mathcal{L}_{\text{min}}$ is the likelihood of the
discarded point being replaced. The sampling
process is performed in the unit hypercube space, with prior
transformations applied to evaluate the likelihood of proposed points in the physical parameter space~\citep{dynesty}. 
The walk length is adaptive on a per-iteration
basis; the algorithm adjusts the number of MCMC steps dynamically to
target a pre-defined number of accepted steps (e.g., 60) for each new
live point generated, up to a maximum limit.

This per-iteration adaptive strategy, however, is ill-suited for GPU
architectures. The variable walk length required for each parallel
sampler would lead to significant thread divergence, where different
cores on the GPU finish their tasks at different times, undermining the
efficiency of the SIMD execution model. To leverage GPU acceleration
effectively, the computational workload must be as uniform as possible
across all parallel processes.

Our implementation therefore preserves the core DE proposal mechanism but
modifies the walk-length adaptation to be compatible with a vectorized
framework. Within the \texttt{blackjax-ns} sampler, the number of MCMC
steps is fixed for all parallel processes within a single batch of
live point updates. The adaptive tuning is then performed at the batch
level. After a batch of $k$ new points has been generated, we compute
the mean acceptance rate across all $k$ walks. The walk length for the
subsequent batch is then adjusted based on this average rate,
using the same logic as \texttt{bilby} to target a desired number of
accepted proposals. Adaptive tuning of walk lengths, or equivalent 
mechanisms~\citep{dau2021wastefreesequentialmontecarlo}, particularly 
in light of the GPU-batched approach, is a key avenue for future work 
to reduce overall runtime.

While this batch-adaptive approach is essential for efficient GPU
vectorization, it introduces some important differences, discussed further in Section~\ref{sec:sampler_config}.
Despite this architectural modification, our kernel is designed to be functionally
analogous to the trusted \texttt{bilby} sampler. This design
represents the most direct translation of the \texttt{bilby} logic
to GPUs, making only the minimal changes required for efficient execution. 

\subsection{Sampler configuration and settings}
\label{sec:sampler_config}

Several key hyperparameters govern the behaviour of the sampler and its
inner kernel. The number of live points, \mbox{$n_{\text{live}}$}, controls the
resolution of the run, where a higher value results in more accurate
posteriors and evidence estimates. The \mbox{$\texttt{n\_accept}$}
parameter sets the target number of accepted steps per MCMC walk,
affecting the decorrelation of samples, and \mbox{$\texttt{maxmcmc}$}
sets the maximum number of attempted steps per walk.

The primary architectural difference between our GPU-based implementation
and the standard CPU-based `acceptance-walk' kernel is the use of
batched sampling. In our framework, a batch of $k$ new points is
generated and added to the live set simultaneously. This batch size is
a user-configurable parameter, \texttt{num\_delete}, and we find that a
value of \mbox{$k \approx 0.5 \times \texttt{n\_live}$} provides a good balance
of parallel efficiency and sampling accuracy for most problems~\citep{yallup2025nested}.


This batched approach has direct consequences for the adaptive tuning of
the MCMC walk length. The tuning is performed only once per \texttt{num\_delete} iterations, 
rather than at every iteration, and every point in a given batch is tuned to have the same walk length.
This design is important for preventing thread divergence and is the 
most natural way to implement this algorithm on a GPU.
However, this less frequent tuning renders the standard \texttt{bilby}
tuning parameter `delay', a smoothing factor for
the adaptation, sub-optimal. 
To achieve a comparable number of accepted steps per
iteration, we therefore adjust this parameter for the parallel
framework. The delay is modified to be a function of the batch size,
recovering the standard \texttt{bilby} behaviour in the sequential
limit, since it is functionally similar to averaging over batches of chains
anyway. 


Another subtle but critical consequence of the architectural shift to GPUs is
its effect on the evolution of the live point set. In the sequential
CPU case, the number of live points is approximately constant throughout the run. In the
batched GPU case, the number of live points follows a `saw-tooth'
pattern, decreasing from $n_{\text{live}}$ to
\mbox{$n_{\text{live}} - k$} over one cycle of $k$ deletions before being replenished (Figure~\ref{fig:nlive_comparison}). This pattern
causes the effective number of live points to be lower than the nominal number,
making it appear that the sampler is able to converge faster than its sequential counterpart
when both are configured with the same number of live points. It is important to 
note that the saw-tooth pattern is common to all
parallel nested sampling, including CPU-based implementations~\citep{Handley:2015vkr},
and it is only the massively parallel nature of the GPU that makes it noticeable.

\begin{figure}
    \centering
    \includegraphics{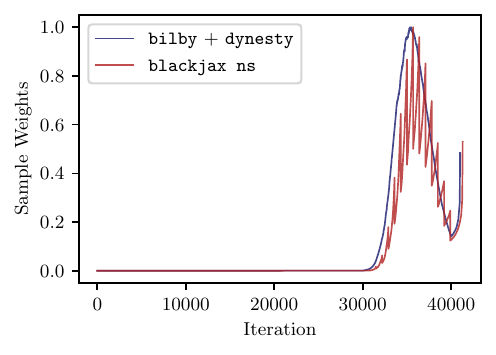}
    \caption{Comparison of the sample weights for each dead point for the sequential CPU and batched GPU implementations.
    The weights are calculated using the prior volumes enclosed between successive dead points and the likelihoods. 
    The shapes are similar, and both implementations enter the bulk of the posterior distribution at similar iterations,
    indiciating that setting the number of live points in \texttt{blackjax-ns} to 1.4 times the number of live points in \texttt{bilby} 
    does indeed result in a like-for-like comparison. The same saw-tooth pattern can be 
    seen in the weights for the \texttt{blackjax-ns} implementation.}
    \label{fig:weights_comparison}
\end{figure}

To conduct a fair and direct comparison between the two frameworks, this
discrepancy must be accounted for. We therefore adjust the number of
live points in the GPU sampler, $n_{\text{GPU}}$, such that the
expected prior volume compression matches that of the CPU sampler. 
Under these settings, both implementations will converge in
approximately the same number of nested sampling iterations (see Figure~\ref{fig:weights_comparison}). 
The following derivation outlines this adjustment.

\subsubsection{Setting $n_{\text{GPU}}$}
\label{sec:setting_n_gpu}

The expected log prior volume fraction, $\log X$, remaining after $k$ iterations
of a nested sampling run is given by~\citep{skilling, dynamic_ns, aeons}
\begin{equation}
    E[\log X_k] = \sum_{i=1}^{k} \frac{1}{n_i},
\end{equation}
where $n_i$ is the number of live points at iteration $i$. In the
CPU-based implementation, $n_i = n_{\text{CPU}}$ is roughly constant.
After $k$ iterations, the expected log volume fraction is therefore
\begin{equation}
    \log X_{\text{CPU}} = \frac{k}{n_{\text{CPU}}}.
\end{equation}
In our GPU implementation, the number of live points decreases over one
batch cycle of $k = \texttt{num\_delete}$ deletions. The total log
volume shrinkage over one such cycle is the sum over the decreasing
number of live points:
\begin{equation}
    \log X_{\text{GPU}} = \sum_{i=0}^{k-1} \frac{1}{n_{\text{GPU}}-i} \approx \ln\left(\frac{n_{\text{GPU}}}{n_{\text{GPU}}-k}\right).
\end{equation}
To ensure a fair comparison, we equate the expected shrinkage of both
methods over one cycle of $k$ iterations:
\begin{equation}
    \frac{k}{n_{\text{CPU}}} \approx \ln\left(\frac{n_{\text{GPU}}}{n_{\text{GPU}}-k}\right).
\end{equation}
Using our recommended setting of $k = 0.5 \times n_{\text{GPU}}$, this
simplifies to $0.5 \times n_{\text{GPU}} / n_{\text{CPU}} \approx \ln(2)$.
We can therefore derive the required number of live points for the GPU
sampler to be
\begin{equation}
    n_{\text{GPU}} \approx 2 \ln(2) \times n_{\text{CPU}} \approx 1.4 \times n_{\text{CPU}}.
\end{equation}
In all comparative analyses presented in this paper, we configure the
number of live points according to this relation to ensure an equal 
effective number of live points and
like-for-like timing comparisons.

\subsection{Likelihood}

To assess the performance of our framework, we employ a standard
frequency-domain likelihood~\citep{lal, NelsonMeyerReview}. The total
speedup in our analysis is achieved by accelerating the two primary
computational components of the inference process: the sampler and the
likelihood evaluation. The former is parallelized at the batch level as
described in Sec.~\ref{sec:methods_kernel}, while the latter is
accelerated using a GPU-native waveform generator.

For this purpose, we generate gravitational waveforms using the
\texttt{ripple} library, which provides GPU-based implementations of
common models~\citep{ripple, wong2023fastgravitationalwaveparameter, Wouters_BNS}. 
This allows the waveform to be calculated
in parallel across the frequency domain and 
across multiple sets of parameters, enabling massive efficiency gains by 
ensuring that this calculation does not become a serial bottleneck.
To isolate the speedup from this combined GPU-based framework,
we deliberately avoid other established acceleration methods like 
heterodyning~\citep{TL_relativebinning, relativebinning2, relativebinning3, relativebinning4}, 
though these are available in the \texttt{jimgw} library too~\citep{wong2023fastgravitationalwaveparameter}.

For the analyses in this paper, we restrict our consideration to binary
black hole systems with aligned spins, for which we use the
\texttt{IMRPhenomD} waveform approximant~\citep{Khan:2015jqa}.
Further details on the specific likelihood configuration for each
analysis, including noise curves and data segments, are provided in Section~\ref{sec:results}.

\subsection{Priors}

For this initial study, we adopt a set of standard, separable priors on
the source parameters, which are summarized in Table~\ref{tab:priors}.
The specific ranges for these priors are dependent on the duration of the signal, and are
also given in Section~\ref{sec:results}.

As is the default within the \texttt{bilby} framework, we sample directly in chirp mass, $\mathcal{M}$, and
mass ratio, $q$. We use priors that are uniform in these parameters directly, instead of uniform in the component masses. 
The aligned spin components, $\chi_1$ and $\chi_2$, are also taken to be uniform over
their allowed range. The coalescence time, $t_c$, is assumed to be uniform
within a narrow window around the signal trigger time.

For the luminosity distance, $d_L$, we adopt a power-law prior of the
form $p(d_L) \propto d_L^2$. This prior corresponds to a distribution of
sources that is uniform in a Euclidean universe. While this is a
simplification that is less accurate at higher redshifts~\citep{bilby_validation}, it is a
standard choice for many analyses and serves as a robust baseline for
this work.

These priors were chosen to facilitate a direct, like-for-like
comparison against the CPU-based \texttt{bilby} and \texttt{dynesty}
framework, and in all such comparisons identical priors were used. The
implementation of more astrophysically motivated, complex prior
distributions for mass, spin, and luminosity distance is left to
future work and is independent of the algorithmic performance.

\begin{table}
\setlength{\tabcolsep}{3pt} 
\centering
\caption{Prior distributions for the parameters of the binary black hole
system. The specific ranges for the masses and spins
 are dependent on the signal and are specified in Section~\ref{sec:results}.}
\label{tab:priors}
\begin{tabular}{l l l c c}
\hline
\hline
\textbf{Parameter} & \textbf{Description} & \textbf{Prior Distribution} & \textbf{Range}\\
\hline
$M_c$ & Chirp Mass & Uniform & - \\
$q$ & Mass Ratio & Uniform & - \\
$\chi_1, \chi_2$ & Aligned spin components & Uniform & - \\
$d_L$ & Luminosity Distance & Power Law (2) & [100, 5000] Mpc \\
$\theta_{\textrm{JN}}$ & Inclination Angle & Sine & [0, $\pi$] rad \\
$\psi$ & Polarization Angle & Uniform & [0, $\pi$] rad \\
$\phi_c$ & Coalescence Phase & Uniform & [0, 2$\pi$] rad \\
$t_c$ & Coalescence Time & Uniform & [-0.1, 0.1] s\\
$\alpha$ & Right Ascension & Uniform & [0, 2$\pi$] rad \\
$\delta$ & Declination & Cosine & [-$\pi$/2, $\pi$/2] rad \\
\hline
\hline
\end{tabular}
\end{table}

\section{Results and discussion}
\label{sec:results}

We now validate the performance of our framework through a series of
analyses. In all of the results, the \texttt{bilby} analyses
were executed on a 16-core CPU instance using Icelake nodes, while the \texttt{blackjax-ns}
analyses were performed on a single NVIDIA L4 GPU, unless otherwise specified.

\subsection{Simulated signals}

\subsubsection{4-second simulated signal}
\label{sec:4s_simulated_signal}

We begin by analysing a 4-second simulated signal from a binary black
hole (BBH) merger. The injection parameters for this signal are
detailed in Table~\ref{tab:injection_params}. To ensure a direct,
like-for-like comparison, the signal was injected into simulated
detector noise using the \texttt{bilby} library, and 
then loaded into both sets of analyses. The analysis
uses a three-detector network, with the design
sensitivity for the fourth LIGO-Virgo-KAGRA observing run (O4).
The frequency range of data analysed is from 20 Hz to 1024 Hz, 
and the network matched filter SNR is 39.6. 

Both the CPU and GPU-based analyses were configured with 1000 effective live
points. As detailed in Sec.~\ref{sec:setting_n_gpu}, this corresponded to
setting the number of live points in \texttt{blackjax-ns} to 1400, with \mbox{\texttt{num\_delete} = 700}.
The termination condition in both cases was set to dlogZ < 0.1 (the default in \texttt{bilby})
and we used the settings \mbox{\texttt{naccept} = 60}, \mbox{\texttt{maxmcmc} = 5000} and \mbox{\texttt{use\_ratio = True}}. 
In both cases, periodic boundary conditions were used for the right ascension, 
polarization angle, and coalescence phase parameters. The prior ranges
for the chirp mass and mass ratio were set to $[25.0, 50.0]~M_{\odot}$
and $[0.25, 1.0]$, respectively, with priors on the aligned
spin components over the range $[-1, 1]$. 

\begin{figure}
    \centering
    \includegraphics{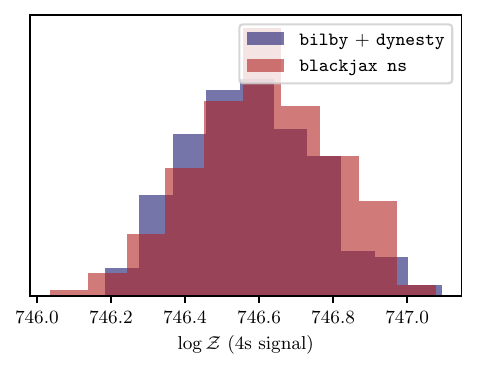}
    \caption{Comparison of the log evidence for the 4s signal. The results are in excellent agreement, demonstrating the
    robustness of the \texttt{blackjax-ns} implementation in recovering the same posteriors and evidence as the \texttt{bilby} implementation.
    This unifies parameter estimation and evidence evaluation into a single GPU-accelerated framework.}
    \label{fig:4s_logZ_comparison}
\end{figure}

The recovered posterior distributions, shown in
Figure~\ref{fig:4s_posteriors}, demonstrate excellent statistical
agreement between the two frameworks. This result validates that our
custom `acceptance-walk' kernel within the vectorized \texttt{blackjax-ns}
framework is functionally equivalent to the trusted sequential
implementation in \texttt{bilby}. The computed log-evidence values are
also in strong agreement, as shown in Figure~\ref{fig:4s_logZ_comparison},
confirming that our implementation provides a robust unified framework for both
parameter estimation and model selection. 

The CPU-based \texttt{bilby} run completed in 2.99 hours on 16 cores, for a total of
47.8 CPU-hours. In contrast, the GPU-based \texttt{blackjax-ns} run
completed in 1.25 hours. This corresponds to a sampling time speedup factor
of 38$\times$. The \texttt{bilby} and \texttt{blackjax-ns} runs performed
62.9 million and 62.5 million likelihood evaluations and had a
mean number of accepted steps per iteration of 28.5 and 30.7, respectively.

Beyond sampling time, we also consider the relative financial cost. Based
on commercial on-demand rates from Google Cloud\footnote{Accessed on 2025-09-03 from https://cloud.google.com/compute/gpus-pricing.}, 
the rental cost for a 16-core
CPU instance and an L4 GPU instance were approximately equivalent at
the time of this work. This equivalence in hourly cost implies a direct
cost-saving factor of approximately 2.4$\times$ for the GPU-based analysis.

In the interest of benchmarking, we also performed the analysis on 
an A100 NVIDIA GPU configured with the same run settings as the L4 GPU.
The A100 is a more powerful GPU and resulted in a lower runtime. Interestingly,
however, when comparing the relative commercial hourly rates of the two GPUs,
the L4 GPU analysis was actually cheaper. Indeed the A100 analysis actually had a 
worse cost compared to the CPU implementation. We summarise these results in 
Table~\ref{tab:4s_time_comparison}.

\begin{table}
    \centering
    \caption{Injection parameters for the 4s signal.}
    \label{tab:injection_params}
    \begin{tabular}{l l l c c}
    \hline
    \hline
    \textbf{Parameter} & \textbf{Value} \\
    \hline
    $\mathcal{M}$ & 35.0 M$_{\odot}$ \\
    $q$ & 0.90 \\
    $\chi_1$ & 0.40 \\
    $\chi_2$ & -0.30 \\
    $d_L$ & 1000 Mpc \\
    $\theta_{\textrm{JN}}$ & 0.40 rad \\
    $\psi$ & 2.66 rad \\
    $\phi$ & 1.30 rad \\
    $t_c$ & 0.0 s\\
    $\alpha$ & 1.38 rad \\
    $\delta$ & -1.21 rad \\
    \hline
    \hline
    \end{tabular}
    \end{table}

\begin{table}
    \setlength{\tabcolsep}{3pt}
    \centering
    \caption{Comparison of the sampling times and cost savings for the 4s signal.}
    \label{tab:4s_time_comparison}
    \begin{tabular}{l l l c c}
    \hline
    \hline
    \textbf{Implementation + Hardware} & \textbf{Time (h)} & \textbf{Speedup} & \textbf{Cost Saving} \\
    \hline
    \texttt{bilby} (16 Icelake CPU cores) & 47.8 & - & - \\
    \texttt{blackjax-ns} (NVIDIA L4) & 1.25 & 38$\times$ & 2.4$\times$ \\
    \texttt{blackjax-ns} (NVIDIA A100) & 0.93 & 51$\times$ & 0.6$\times$ \\
    \hline
    \hline
    \end{tabular}
    \end{table}

\subsection{Injection Study}
\label{sec:injection_study}

\begin{figure}
    \centering
    \includegraphics{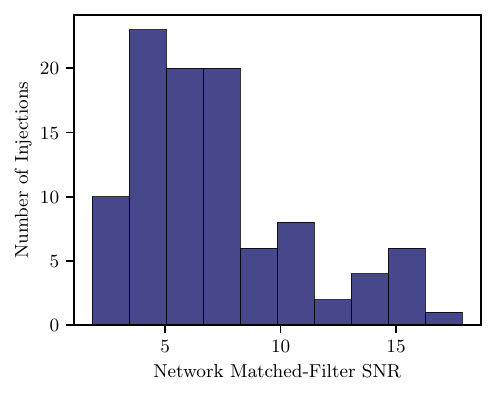}
    \caption{Distribution of the network signal-to-noise ratios (SNR) for the injected signals.}
    \label{fig:snr_dist}
\end{figure}

To systematically assess the performance and robustness of our framework
across a diverse parameter space, we performed an injection study
comparing our GPU-based \texttt{blackjax-ns} sampler against the
CPU-based \texttt{bilby}+\texttt{dynesty} implementation. A population
of 100 simulated BBH signals was generated using \texttt{bilby} and
injected into noise representative of the O4 detector network
sensitivity. The network signal-to-noise ratios (SNR) for this
injection set span a range from 1.84 to 17.87 (Figure~\ref{fig:snr_dist}).
As above, we use a three-detector network for the analysis.

For this study, the prior ranges were set to $[20.0, 50.0]~M_{\odot}$
for the chirp mass and $[0.5, 1.0]$ for the mass ratio. The aligned
spin components were bounded by uniform priors over the range
$[-0.8, 0.8]$. All other prior distributions are as defined in
Table~\ref{tab:priors}.

All analyses in this study were performed using 1000 live points for \texttt{bilby} and
1400 live points for \texttt{blackjax-ns}. Given that quite a few signals in the set have a low SNR, and therefore a
low Bayes factor when comparing the signal hypothesis to the noise-only
hypothesis, a more robust termination condition was required to
ensure accurate evidence computation as well as posterior estimation.
We therefore set the termination criterion based on the fractional
remaining evidence, such that the analysis stops when the estimated
evidence in the live points is less than 0.1\% of the accumulated
evidence. Both samplers were configured with
\mbox{\texttt{naccept} = 60} and \mbox{\texttt{maxmcmc} = 5000}, and the \texttt{blackjax-ns}
sampler used a batch size of \mbox{\texttt{num\_delete} = 700}.

\begin{figure}
    \centering
    \includegraphics{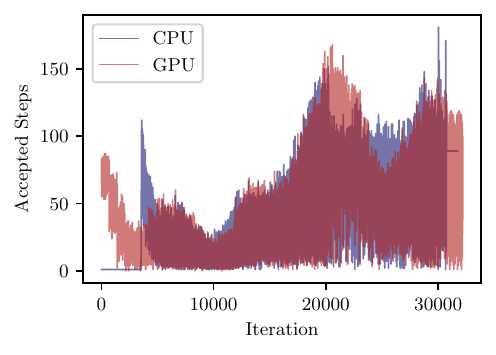}
    \caption{Comparison of the accepted number of steps per iteration for the sequential CPU and batched GPU analyses of
    the first signal from the injection study.
    We adapt our `delay' parameter from the tuning formula such that we obtain similar accepted steps for the 
    two implementations. The \texttt{blackjax-ns} implementation can only perform tuning every \mbox{\texttt{num\_delete}} iterations,
    so we tune the chain length more aggressively.}
    \label{fig:chain_length_comparison}
\end{figure}

\begin{figure}
    \centering
    \includegraphics[width=\columnwidth]{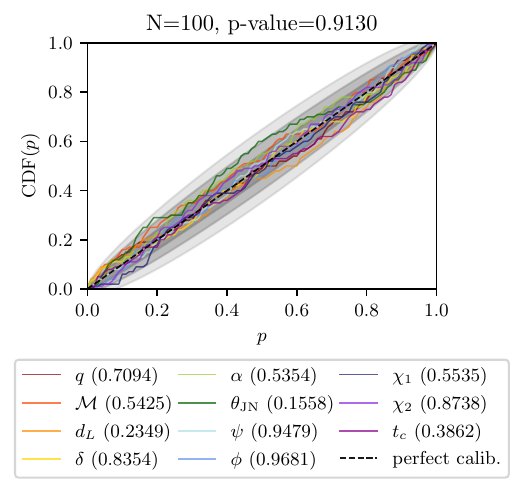}
    \caption{Percentile-percentile (PP) coverage plot for the
    100-injection study, obtained with the \texttt{blackjax-ns}
    sampler. The cumulative fraction of events where the true
    injected parameter falls below a given credible level is plotted
    against that credible level. The proximity of all parameter curves
    to the diagonal indicates that the posterior credible intervals
    are statistically well-calibrated. A corresponding plot for the
    \texttt{bilby+dynesty} analysis, which should be identical,
    is provided in the Appendix for reference.}
    \label{fig:pp_coverage}
\end{figure}

\begin{figure}
    \centering
    \includegraphics{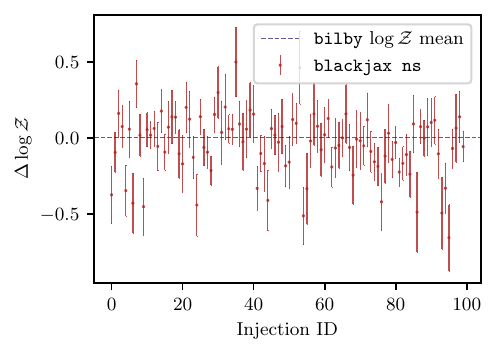}
    \caption{The log evidence difference between the 
    \texttt{blackjax-ns} and \texttt{bilby+dynesty} 
    samplers for each injection in the study. For each estimate,
    we take the mean log evidence estimated by \texttt{bilby} to be
    the reference value, and subtract the \texttt{bilby} estimates from the 
    log evidences calculated by \texttt{blackjax-ns}.
    The error bars represent 1$\sigma$ confidence intervals for these differences.
    There appears to be some underestimation of the error bars, potentially due
    to some sets of samples not being as well decorrelated as others, 
    despite a reasonable average decorrelation. However, a calibration test
    was performed to quantify this. For the $\Delta \log \mathcal{Z}$
    estimates, we calculate the percentiles at which $\Delta \log \mathcal{Z} = 0$
    lies. If the error bars are correctly calibrated, the resulting distribution of
    percentiles should be uniform; the Kolmogorov-Smirnov test returns a p-value of 
    $0.062$, indicating that at the $5\%$ significance level the error bars are 
    well-calibrated.}
    \label{fig:evidence_vs_steps_iters}
\end{figure}

The resulting PP plot for our \texttt{blackjax-ns} sampler
with the `acceptance-walk' kernel is shown in Figure~\ref{fig:pp_coverage}.
This plot evaluates the self-consistency of the posteriors by checking
the distribution of true injected parameter values within their
recovered credible intervals. The results demonstrate that the credible
intervals are well-calibrated, with the cumulative distributions for all
parameters lying within the expected statistical variance of the
line of perfect calibration. This confirms the robustness of our implementation. The full
set of posterior and evidence results for all 100 injections is made
publicly available in the data repository accompanying this paper~\citep{Prathaban_2025_zenodo}.
We show the consistency of the evidence estimates in Figure~\ref{fig:evidence_vs_steps_iters}.

\begin{figure}
    \centering
    \includegraphics{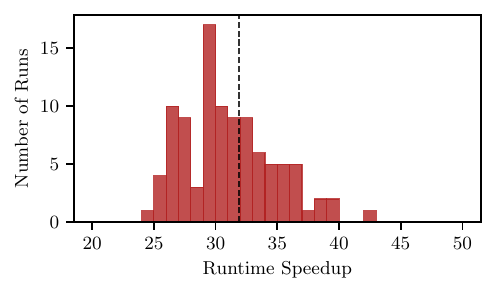}
    \caption{The runtime speedups for all 100 events in the injection study. For the
    CPU-based analyses, we take the wall-time and multiply by 16 to get the total runtime. 
    For the GPU-based analyses, we use the wall-time directly.}
    \label{fig:speedup_comparison}
\end{figure}

\begin{figure}
    \centering
    \includegraphics{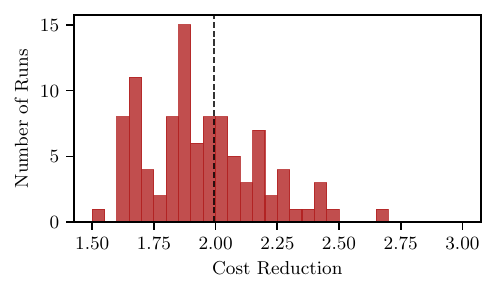}
    \caption{The cost reductions for all 100 events in the injection study.
    Although the runtime speedups are significant, the more honest and fair comparison
    is of the relative costs of the two sets of analyses, since GPUs are more
    expensive than CPUs. We use the commercial on-demand rates for a 16-core CPU
    instance and an L4 GPU instance at the time of this work, and compare the cost savings
    obtained from using our GPU-accelerated pipeline.}
    \label{fig:cost_reduction}
\end{figure}

The average number of likelihood evaluations for the \texttt{blackjax-ns}
sampler was 38.4 million, consistent with the average of 37.5 million for the
\texttt{bilby+dynesty} sampler. A more detailed breakdown of all run
statistics is provided in Appendix~\ref{app:injection_study_stats}.
In terms of runtime, the \texttt{blackjax-ns} sampler completed the analyses
in an average of 0.82 hours per injection, with individual run
times ranging from 32 to 76 minutes.
In contrast, the CPU-based runs
required an average of 26.3 total CPU-hours, with some runs taking as long as 
45 hours. 
Figure~\ref{fig:speedup_comparison} shows the distribution of the
resulting runtime speedup factors, which have a mean value of
32$\times$. As in 
section~\ref{sec:4s_simulated_signal}, we also translate this performance 
gain into a cost reduction using commercial cloud computing rates.
As shown in Figure~\ref{fig:cost_reduction}, the speedup
corresponds to an average cost-reduction factor of 2.0$\times$ for the
GPU-based analysis compared to its CPU-based counterpart.

\subsection{Disentangling Sources of GPU Acceleration}
\label{sec:disentangling_acceleration}

The overall performance gain of our framework arises from two distinct
forms of parallelisation: the parallel evaluation of a single
likelihood across its frequency bins (intra-likelihood), and the
parallel sampling process itself, where multiple MCMC chains are run
simultaneously (inter-sample). In this section, we attempt to disentangle
these two contributions.

We configured our \texttt{blackjax-ns} sampler to run in a sequential
mode. This was achieved by setting the batch size to one
(\texttt{num\_delete} = 1) and the number of live points to 1000,
identical to the CPU analysis. With a batch size of one, the
`saw-tooth' pattern in the live point population described in
Sec.~\ref{sec:sampler_config} is eliminated, making the effective
number of live points equal to the nominal value. This removes the need
for the corrective scaling of $n_{\text{live}}$ applied in our main
parallel analyses, and we can also use the same `delay' for tuning the walk lengths
as for the \texttt{bilby} runs. In this configuration, the algorithm proceeds
analogously to the \texttt{bilby} implementation, and the only source of
acceleration relative to the CPU is the GPU's ability to parallelise the
likelihood calculation over the frequency domain.

For this test, we used the first signal from our injection study (network
SNR of 8.82). The baseline CPU-based \texttt{bilby} analysis required
30.4 total CPU-hours to complete. The sequential-GPU analysis, which
benefits only from intra-likelihood parallelisation, completed in
9.1 hours. This represents a speedup of 3.3$\times$. Finally, the
fully parallel run, using its correctly scaled live point count of 1400 and
a batch size of \texttt{num\_delete} = 700, completed in 0.82 hours.
This corresponds to a further speedup of 11.1$\times$ over the
sequential-GPU run.

This result demonstrates that while a GPU-native likelihood provides
a significant performance benefit, the more dominant contribution to the
overall speedup (a total of 37.1$\times$ in this case) originates from the
massively parallel, batched sampling architecture. This result
underscores the importance of the algorithmic reformulation 
of nested sampling pioneered in~\cite{yallup2025nested}.

\subsection{8-second simulated signal}
\label{sec:8s_simulated_signal}

To investigate the scalability of our framework with a more
computationally demanding likelihood, we now analyse a simulated
8-second BBH signal. The injection parameters are detailed in
Table~\ref{tab:8s_injection_params}. For this analysis, the data are
evaluated between 20 Hz and 2048 Hz, quadrupling the number of
frequency bins compared to the 4-second signal analysis. As above, the signal was
injected into noise from a three-detector network with O4 sensitivity,
resulting in a network SNR of 11.25.

The prior ranges for the chirp mass and mass ratio were set to
$[10.0, 25.0]~M_{\odot}$ and $[0.25, 1.0]$, respectively. The aligned
spin components were bounded by uniform priors over the range
$[-0.8, 0.8]$, with all other prior distributions as defined in
Table~\ref{tab:priors}. The sampler configurations were identical to
those used in the injection study (Sec.~\ref{sec:injection_study}), with
$n_{\text{live}}=1400$ for \texttt{blackjax-ns} and
$n_{\text{live}}=1000$ for \texttt{bilby}, and an \texttt{naccept}
target of 60. Again, we use the 
termination condition described in Section~\ref{sec:injection_study}.

\begin{figure*}
    \centering
    \includegraphics{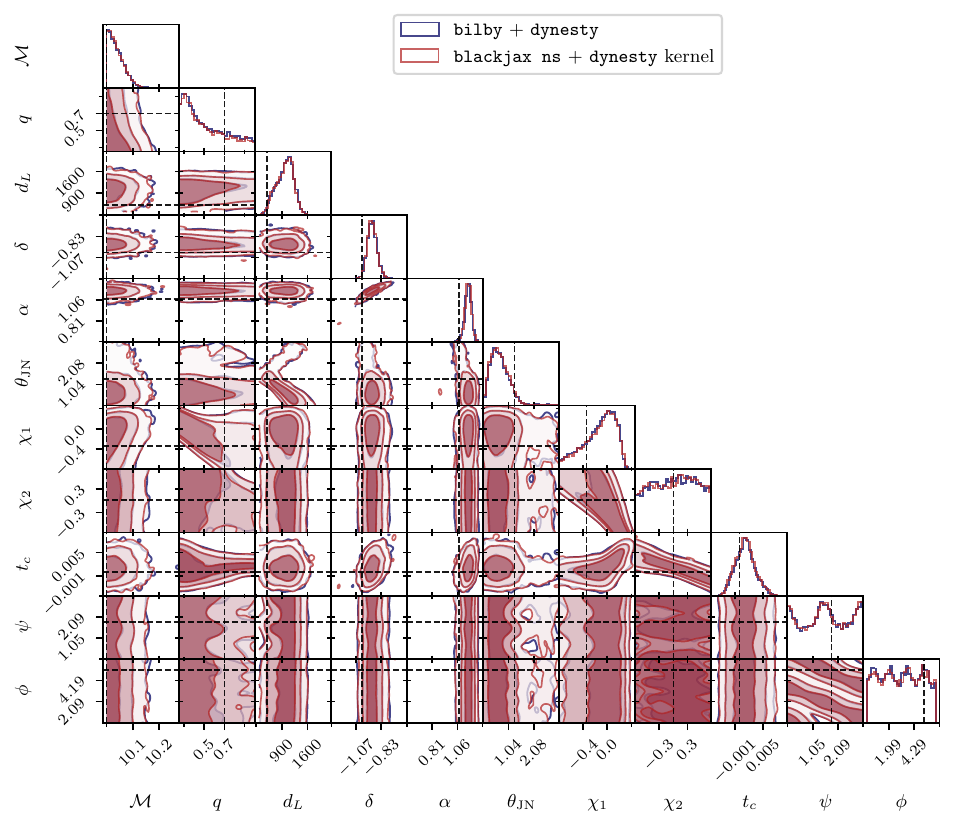}
    \caption{Recovered posterior distributions for the 8s simulated
    signal, comparing our GPU-based \texttt{blackjax-ns} sampler
    with the CPU-based \texttt{bilby} sampler. The injected values
    are marked by black lines. The strong statistical agreement confirms
    the validity of our implementation for longer-duration signals.
    The GPU implementation provided a wall-time speedup of 46$\times$ and a cost reduction of 2.9$\times$.}
    \label{fig:8s_posteriors}
\end{figure*}

\begin{figure}
    \centering
    \includegraphics{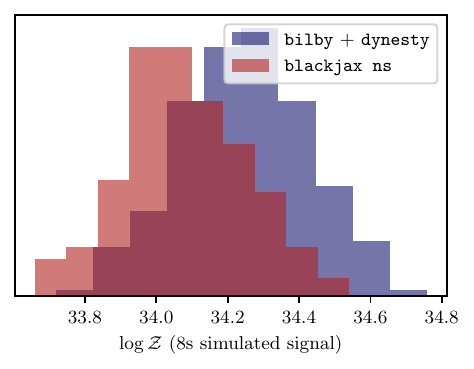}
    \caption{Comparison of the recovered log-evidence ($Z$) for the 8s
    signal. The results from both the \texttt{bilby} and
    \texttt{blackjax-ns} frameworks are consistent within their
    estimated numerical uncertainties.}
    \label{fig:8s_logZ}
\end{figure}

The recovered posterior distributions and log-evidence values are shown
in Figure~\ref{fig:8s_posteriors} and Figure~\ref{fig:8s_logZ}. We again
find excellent agreement between the two frameworks, further validating
the accuracy of our implementation, despite the differences arising from 
parallelisation. The CPU-based \texttt{bilby} analysis required 167.2 total CPU-hours to
converge, performing 58 million likelihood evaluations. Our GPU-based
\texttt{blackjax-ns} sampler completed the same analysis in 4.55 hours, with 61 million likelihood evaluations,
corresponding to a wall-time speedup factor of 37$\times$. 
The associated cost-reduction factor for this
analysis was 2.3$\times$. 

The scaling of the runtime for this longer-duration signal provides a
key insight into the practical limits of GPU parallelisation. In an
ideal model where the GPU has sufficient parallel cores for every
frequency bin, the likelihood evaluation time would be independent of
signal duration. In such a scenario, the total runtime would be dictated
primarily by the number of likelihood evaluations needed for convergence.
In this case, the analysis should have taken roughly the same
time as the 4-second signal analysis, or $1.25 \times 61/63 = 1.21$ hours. 
However, we observe that the runtime increased significantly more than
what would be predicted by the relative number of likelihood evaluations alone.

This discrepancy arises because the computational load of the larger
frequency array exceeds the parallel processing capacity of the L4 GPU.
As a result, the GPU's internal scheduler must batch the calculation
across the frequency domain, re-introducing a partial serial dependency
that makes the likelihood evaluation time scale with the number of bins.
This result serves as an important practical clarification to a common
argument for GPU acceleration. While the idealised model of
parallelisation suggests that evaluation time should be independent of
signal duration, our work demonstrates that this does not hold
indefinitely. Once the problem size saturates the GPU's finite
resources, such as its compute or memory bandwidth, the runtime once
again begins to scale with the number of frequency bins, a behaviour
analogous to the scaling seen in the CPU-based case.


\section{Conclusions}
\label{sec:conclusions}

\begin{table}
    \centering
    \caption{Injection parameters for the 8s signal.}
    \label{tab:8s_injection_params}
    \begin{tabular}{l l l c c}
    \hline
    \hline
    \textbf{Parameter} & \textbf{Value} \\
    \hline
    $\mathcal{M}$ & 10.0 M$_{\odot}$ \\
    $q$ & 0.70 \\
    $\chi_1$ & -0.34 \\
    $\chi_2$ & 0.01 \\
    $d_L$ & 500 Mpc \\
    $\theta_{\textrm{JN}}$ & 1.30 rad \\
    $\psi$ & 1.83 rad \\
    $\phi$ & 5.21 rad \\
    $t_c$ & 0.0 s\\
    $\alpha$ & 1.07 rad \\
    $\delta$ & -1.01 rad \\
    \hline
    \hline
    \end{tabular}
    \end{table}

In this work, we have presented the development and validation of a
GPU-accelerated implementation of the `acceptance-walk' nested sampling
kernel, a trusted algorithm from the community-standard \texttt{bilby}
and \texttt{dynesty} framework. By integrating this kernel into the
vectorized \texttt{blackjax-ns} framework, we have created a tool for
rapid, unified Bayesian parameter estimation and model selection.
Through systematic studies of simulated binary black hole signals, we
demonstrated that our implementation is functionally analogous to its
CPU-based counterpart, producing statistically consistent posterior
distributions and evidence estimates with typical wall-time speedups of
20-40$\times$ and cost reductions of 1.5-2.5$\times$.

It is important to contextualize the cost savings reported in this
work. As the trajectory of
high-performance computing is increasingly driven by large-scale
investment in GPU hardware for artificial intelligence, we
anticipate that the cost-per-computation on these devices will continue
to fall relative to CPUs. Therefore, the cost-saving advantage of our
approach is expected to become more pronounced in the future, even
if the relative runtimes remain unchanged. 
Furthermore, the JAX~\citep{jax2018github} ecosystem is very well 
suited to distribution over multiple devices and 
architectures. It has been closely developed alongside the 
Tensor Processing Unit (TPU) architecture~\citep{jouppi2023tpu}, 
offering a clear avenue to integrate this work with such 
frontier low-cost hardware.

A key conclusion of this work is that the performance gains 
reported here are primarily a consequence
of the architectural shift to the GPU. This implies that similar relative
speedups are achievable for more computationally demanding waveforms,
such as those including spin precession and higher order modes (e.g., 
\texttt{IMRPhenomXPHM}~\citep{IMRPhenomXPHM}), provided they are also
implemented in a GPU-native framework. Our analysis of a longer-duration
signal confirmed, however, that this scalability is ultimately limited
by the finite resources of the hardware, a practical constraint that can
be addressed with more powerful GPUs or likelihood-acceleration techniques
like heterodyning~\citep{TL_relativebinning,relativebinning2,relativebinning3,relativebinning4}.
Importantly, by significantly reducing the computational cost
of nested sampling, our framework makes tackling previously prohibitive
next-generation analyses~\citep{HuAccelerationReview, makaibaker2025}, 
such as those involving long-duration
signals or high SNR signals, a more viable prospect.

Beyond future-proofing the community's core analysis
tools, this work establishes an important performance baseline. 
As the community develops novel, GPU-accelerated
sampling algorithms, it is essential to disentangle performance gains
originating from the hardware parallelization itself from those arising
from genuine algorithmic innovation. By developing a functionally
equivalent, GPU-native version of one of the community-standard 
algorithms, we have isolated and quantified the speedup attributable
solely to the architectural shift, which allows for batched sampling and
parallelized likelihood evaluations on a GPU.
This provides a robust reference against which future, more advanced
GPU-based samplers can be rigorously benchmarked, in order to 
guide the development of next generation Bayesian inference tools.

There are several future avenues of research and development. 
Having established this performance baseline, we can now rigorously
evaluate novel inner sampling kernels designed specifically for the GPU
architecture. While the speedups reported in this work are significant,
the `acceptance-walk' sampling method was never designed to be 
optimised for the GPU architecture. This motivates the exploration of 
alternative inner sampling kernels, such as those based on the 
Hit-and-Run Slice Sampling (HRSS) algorithm introduced in~\cite{yallup2025nested}, 
which may offer more efficient performance on this architecture.

GPU-accelerated sampling algorithms such as the one presented in this work can only run when
paired with a GPU-native likelihood. Our work therefore provides additional 
motivation for the continued development
and porting of more complex and physically comprehensive waveform models
to GPU-based libraries like \texttt{ripple}~\citep{ripple}. On the software side, future
work will involve integrating additional features from \texttt{bilby}~\citep{bilby_paper}
into our framework, such as the astrophysically motivated prior
distributions for masses, spins, and luminosity distance, to enable 
more realistic analyses that are of interest to the GW community.

Finally, the batched architecture of our sampler is highly compatible
with machine learning techniques used for accelerating Bayesian
inference. Methods that use normalizing flows to speed up
the core nested sampling algorithm, such as those in~\cite{Williams2021Nessai}
and~\cite{Prathaban}, are most
computationally efficient when they can evaluate a large number of
samples in parallel. The serial nature of a conventional CPU-based
sampler represents a significant bottleneck for these models. In
contrast, our parallel framework is a natural fit for these operational 
requirements. This compatibility enables the development of highly efficient,
hybrid inference pipelines that combine GPU-accelerated sampling with
GPU-native machine learning models, further lowering the computational
cost of future gravitational-wave analyses.

\section*{Acknowledgements}

This work was supported by the research environment and infrastructure of the Handley Lab at the University of Cambridge.
MP is supported by the Harding Distinguished Postgraduate Scholars Programme (HDPSP). 
JA is supported by a fellowship from the Kavli Foundation.
This work was performed using the Cambridge Service for Data Driven Discovery (CSD3), 
part of which is operated by the University of Cambridge Research Computing on behalf of the
STFC DiRAC HPC Facility (www.dirac.ac.uk). The DiRAC component of CSD3 was funded by BEIS 
capital funding via STFC capital grants ST/P002307/1 and ST/R002452/1 and STFC operations 
grant ST/R00689X/1. DiRAC is part of the National e-Infrastructure.
This material is based upon work supported by the Google Cloud research
credits program, with the award GCP397499138.

\section*{Data Availability}

All the data used in this analysis, including the relevant nested sampling chains, 
can be obtained from~\cite{Prathaban_2025_zenodo}. We include a file with all the 
code to reproduce the plots in this paper. The \texttt{blackjax-ns} code is available 
from~\cite{blackjax_ns_github}, and the custom kernel implementated in this paper can
be found at~\cite{Prathaban_2025_github}. The latter link also contains instructions and
example files on how to use the kernel to run GW analyses.

\section*{Software and Tools}

The authors acknowledge the use of AI tools in preparing this manuscript 
and the accompanying software. Generative models, specifically Gemini 
2.5 Pro and Claude 4.0 Sonnet, were used to assist with drafting and 
refining the text. Additionally, the AI coding agent Claude Code 
was utilized during software development. All AI-generated 
outputs, including text and code, were carefully reviewed, 
edited, and validated by the authors to ensure their accuracy.



\bibliographystyle{mnras}
\bibliography{references} 

\begin{thebibliography}{}
\makeatletter
\relax
\def\mn@urlcharsother{\let\do\@makeother \do\$\do\&\do\#\do\^\do\_\do\%\do\~}
\def\mn@doi{\begingroup\mn@urlcharsother \@ifnextchar [ {\mn@doi@} {\mn@doi@[]}}
\def\mn@doi@[#1]#2{\def\@tempa{#1}\ifx\@tempa\@empty \href {http://dx.doi.org/#2} {doi:#2}\else \href {http://dx.doi.org/#2} {#1}\fi \endgroup}
\def\mn@eprint#1#2{\mn@eprint@#1:#2::\@nil}
\def\mn@eprint@arXiv#1{\href {http://arxiv.org/abs/#1} {{\tt arXiv:#1}}}
\def\mn@eprint@dblp#1{\href {http://dblp.uni-trier.de/rec/bibtex/#1.xml} {dblp:#1}}
\def\mn@eprint@#1:#2:#3:#4\@nil{\def\@tempa {#1}\def\@tempb {#2}\def\@tempc {#3}\ifx \@tempc \@empty \let \@tempc \@tempb \let \@tempb \@tempa \fi \ifx \@tempb \@empty \def\@tempb {arXiv}\fi \@ifundefined {mn@eprint@\@tempb}{\@tempb:\@tempc}{\expandafter \expandafter \csname mn@eprint@\@tempb\endcsname \expandafter{\@tempc}}}

\bibitem[\protect\citeauthoryear{Abbott et~al.}{Abbott et~al.}{2016}]{GW150914}
Abbott B.~P.,  et~al., 2016, \mn@doi [Phys. Rev. Lett.] {10.1103/PhysRevLett.116.061102}, 116, 061102

\bibitem[\protect\citeauthoryear{Abbott et~al.}{Abbott et~al.}{2017a}]{GW170817}
Abbott B.~P.,  et~al., 2017a, \mn@doi [Phys. Rev. Lett.] {10.1103/PhysRevLett.119.161101}, 119, 161101

\bibitem[\protect\citeauthoryear{Abbott et~al.}{Abbott et~al.}{2017b}]{siren}
Abbott B.~P.,  et~al., 2017b, \mn@doi [Nature] {10.1038/nature24471}, 551, 85

\bibitem[\protect\citeauthoryear{Abbott et~al.}{Abbott et~al.}{2019}]{GWTC1}
Abbott B.~P.,  et~al., 2019, \mn@doi [Phys. Rev. X] {10.1103/PhysRevX.9.031040}, 9, 031040

\bibitem[\protect\citeauthoryear{Abbott et~al.}{Abbott et~al.}{2020a}]{aLVK_prospects}
Abbott B.~P.,  et~al., 2020a, \mn@doi [Living Rev. Rel.] {10.1007/s41114-020-00026-9}, 23, 3

\bibitem[\protect\citeauthoryear{Abbott et~al.}{Abbott et~al.}{2020b}]{LIGO_guide_signalextraction}
Abbott B.~P.,  et~al., 2020b, \mn@doi [Classical and Quantum Gravity] {10.1088/1361-6382/ab685e}, 37, 055002

\bibitem[\protect\citeauthoryear{Abbott et~al.}{Abbott et~al.}{2021}]{GWTC2_GR}
Abbott R.,  et~al., 2021, \mn@doi [Phys. Rev. D] {10.1103/PhysRevD.103.122002}, 103, 122002

\bibitem[\protect\citeauthoryear{Abbott et~al.}{Abbott et~al.}{2023a}]{GWTC3_pop_analysis}
Abbott R.,  et~al., 2023a, \mn@doi [Phys. Rev. X] {10.1103/PhysRevX.13.011048}, 13, 011048

\bibitem[\protect\citeauthoryear{Abbott et~al.}{Abbott et~al.}{2023b}]{GWTC3}
Abbott R.,  et~al., 2023b, \mn@doi [Phys. Rev. X] {10.1103/PhysRevX.13.041039}, 13, 041039

\bibitem[\protect\citeauthoryear{Abbott et~al.}{Abbott et~al.}{2024}]{GWTC2}
Abbott R.,  et~al., 2024, \mn@doi [Phys. Rev. D] {10.1103/PhysRevD.109.022001}, 109, 022001

\bibitem[\protect\citeauthoryear{Acernese et~al.,}{Acernese et~al.}{2014}]{aVirgo}
Acernese F.,  et~al., 2014, Classical and Quantum Gravity, 32, 024001

\bibitem[\protect\citeauthoryear{Ashton et~al.}{Ashton et~al.}{2019}]{bilby_paper}
Ashton G.,  et~al., 2019, \mn@doi [Astrophys. J. Suppl.] {10.3847/1538-4365/ab06fc}, 241, 27

\bibitem[\protect\citeauthoryear{Ashton, Bernstein, Buchner  \& et al.}{Ashton et~al.}{2022}]{NSNature}
Ashton G.,  Bernstein N.,  Buchner J.,   et al. 2022, \mn@doi [Nature Reviews Methods Primers] {10.1038/s43586-022-00121-x}, 2, 39

\bibitem[\protect\citeauthoryear{Baker, Lasky, Thrane  \& Golomb}{Baker et~al.}{2025}]{makaibaker2025}
Baker A.~M.,  Lasky P.~D.,  Thrane E.,   Golomb J.,  2025, Significant challenges for astrophysical inference with next-generation gravitational-wave observatories (\mn@eprint {arXiv} {2503.04073})

\bibitem[\protect\citeauthoryear{{Bayes}}{{Bayes}}{1763}]{Bayes1763}
{Bayes} T.,  1763, \mn@doi [Philosophical Transactions of the Royal Society of London] {10.1098/rstl.1763.0053}, \href {https://ui.adsabs.harvard.edu/abs/1763RSPT...53..370B} {53, 370}

\bibitem[\protect\citeauthoryear{{Biwer}, {Capano}, {De}, {Cabero}, {Brown}, {Nitz}  \& {Raymond}}{{Biwer} et~al.}{2019}]{pycbc}
{Biwer} C.~M.,  {Capano} C.~D.,  {De} S.,  {Cabero} M.,  {Brown} D.~A.,  {Nitz} A.~H.,   {Raymond} V.,  2019, \mn@doi [Publications of the Astronomical Society of the Pacific] {10.1088/1538-3873/aaef0b}, \href {https://ui.adsabs.harvard.edu/abs/2019PASP..131b4503B} {131, 024503}

\bibitem[\protect\citeauthoryear{Bradbury et~al.,}{Bradbury et~al.}{2018}]{jax2018github}
Bradbury J.,  et~al., 2018, {JAX}: composable transformations of {P}ython+{N}um{P}y programs, \url {http://github.com/google/jax}

\bibitem[\protect\citeauthoryear{Branchesi et~al.}{Branchesi et~al.}{2023}]{ET_science_case}
Branchesi M.,  et~al., 2023, \mn@doi [Journal of Cosmology and Astroparticle Physics] {10.1088/1475-7516/2023/07/068}, 2023, 068

\bibitem[\protect\citeauthoryear{Buchner}{Buchner}{2023}]{NS_methods_buchner}
Buchner J.,  2023, \mn@doi [Statistics Surveys] {10.1214/23-SS144}, 17, 169

\bibitem[\protect\citeauthoryear{Cabezas, Corenflos, Lao  \& Louf}{Cabezas et~al.}{2024}]{cabezas2024blackjax}
Cabezas A.,  Corenflos A.,  Lao J.,   Louf R.,  2024, BlackJAX: Composable {B}ayesian inference in {JAX} (\mn@eprint {arXiv} {2402.10797})

\bibitem[\protect\citeauthoryear{Christensen \& Meyer}{Christensen \& Meyer}{2022}]{NelsonMeyerReview}
Christensen N.,  Meyer R.,  2022, \mn@doi [Rev. Mod. Phys.] {10.1103/RevModPhys.94.025001}, 94, 025001

\bibitem[\protect\citeauthoryear{Cornish}{Cornish}{2013}]{relativebinning4}
Cornish N.~J.,  2013, Fast Fisher Matrices and Lazy Likelihoods (\mn@eprint {arXiv} {1007.4820}), \url {https://arxiv.org/abs/1007.4820}

\bibitem[\protect\citeauthoryear{Dau \& Chopin}{Dau \& Chopin}{2021}]{dau2021wastefreesequentialmontecarlo}
Dau H.-D.,  Chopin N.,  2021, Waste-free Sequential Monte Carlo (\mn@eprint {arXiv} {2011.02328}), \url {https://arxiv.org/abs/2011.02328}

\bibitem[\protect\citeauthoryear{Edwards, Wong, Lam, Coogan, Foreman-Mackey, Isi  \& Zimmerman}{Edwards et~al.}{2024}]{ripple}
Edwards T. D.~P.,  Wong K. W.~K.,  Lam K. K.~H.,  Coogan A.,  Foreman-Mackey D.,  Isi M.,   Zimmerman A.,  2024, \mn@doi [Phys. Rev. D] {10.1103/PhysRevD.110.064028}, 110, 064028

\bibitem[\protect\citeauthoryear{Handley, Hobson  \& Lasenby}{Handley et~al.}{2015}]{Handley:2015vkr}
Handley W.~J.,  Hobson M.~P.,   Lasenby A.~N.,  2015, \mn@doi [Mon. Not. Roy. Astron. Soc.] {10.1093/mnras/stv1911}, 453, 4385

\bibitem[\protect\citeauthoryear{Henderson \& Goggans}{Henderson \& Goggans}{2014}]{parallel_ns}
Henderson R.~W.,  Goggans P.~M.,  2014, in AIP Conference Proceedings. pp 100--105, \mn@doi{10.1063/1.4903717}

\bibitem[\protect\citeauthoryear{Higson, Handley, Hobson  \& Lasenby}{Higson et~al.}{2018}]{Higson_Errors}
Higson E.,  Handley W.,  Hobson M.,   Lasenby A.,  2018, \mn@doi [Bayesian Analysis] {10.1214/17-BA1075}, 13, 873

\bibitem[\protect\citeauthoryear{Higson, Handley, Hobson  \& Lasenby}{Higson et~al.}{2019}]{dynamic_ns}
Higson E.,  Handley W.,  Hobson M.,   Lasenby A.,  2019, \mn@doi [Statistics and Computing] {10.1007/s11222-018-9844-0}, 29, 891

\bibitem[\protect\citeauthoryear{Hoffman \& Sountsov}{Hoffman \& Sountsov}{2022}]{pmlr-v151-hoffman22a}
Hoffman M.~D.,  Sountsov P.,  2022, in Camps-Valls G.,  Ruiz F. J.~R.,   Valera I.,  eds,  Proceedings of Machine Learning Research Vol. 151, Proceedings of The 25th International Conference on Artificial Intelligence and Statistics. PMLR, pp 7799--7813, \url {https://proceedings.mlr.press/v151/hoffman22a.html}

\bibitem[\protect\citeauthoryear{Hoffman, Radul  \& Sountsov}{Hoffman et~al.}{2021}]{pmlr-v130-hoffman21a}
Hoffman M.,  Radul A.,   Sountsov P.,  2021, in Banerjee A.,  Fukumizu K.,  eds,  Proceedings of Machine Learning Research Vol. 130, Proceedings of The 24th International Conference on Artificial Intelligence and Statistics. PMLR, pp 3907--3915, \url {https://proceedings.mlr.press/v130/hoffman21a.html}

\bibitem[\protect\citeauthoryear{Hu \& Veitch}{Hu \& Veitch}{2024}]{HuAccelerationReview}
Hu Q.,  Veitch J.,  2024, Costs of Bayesian Parameter Estimation in Third-Generation Gravitational Wave Detectors: a Review of Acceleration Methods (\mn@eprint {arXiv} {2412.02651}), \url {https://arxiv.org/abs/2412.02651}

\bibitem[\protect\citeauthoryear{Hu, Baryshnikov  \& Handley}{Hu et~al.}{2024}]{aeons}
Hu Z.,  Baryshnikov A.,   Handley W.,  2024, \mn@doi [Mon. Not. Roy. Astron. Soc.] {10.1093/mnras/stae1754}, 532, 4035

\bibitem[\protect\citeauthoryear{Jouppi et~al.,}{Jouppi et~al.}{2023}]{jouppi2023tpu}
Jouppi N.,  et~al., 2023, in Proceedings of the 50th annual international symposium on computer architecture. pp 1--14

\bibitem[\protect\citeauthoryear{Khan, Husa, Hannam, Ohme, P\"urrer, Jim\'enez~Forteza  \& Boh\'e}{Khan et~al.}{2016}]{Khan:2015jqa}
Khan S.,  Husa S.,  Hannam M.,  Ohme F.,  P\"urrer M.,  Jim\'enez~Forteza X.,   Boh\'e A.,  2016, \mn@doi [Phys. Rev. D] {10.1103/PhysRevD.93.044007}, 93, 044007

\bibitem[\protect\citeauthoryear{Krishna, Vijaykumar, Ganguly, Talbot, Biscoveanu, George, Williams  \& Zimmerman}{Krishna et~al.}{2023}]{TL_relativebinning}
Krishna K.,  Vijaykumar A.,  Ganguly A.,  Talbot C.,  Biscoveanu S.,  George R.~N.,  Williams N.,   Zimmerman A.,  2023, Accelerated parameter estimation in Bilby with relative binning (\mn@eprint {arXiv} {2312.06009}), \url {https://arxiv.org/abs/2312.06009}

\bibitem[\protect\citeauthoryear{Leslie, Dai  \& Pratten}{Leslie et~al.}{2021}]{relativebinning3}
Leslie N.,  Dai L.,   Pratten G.,  2021, \mn@doi [Physical Review D] {10.1103/physrevd.104.123030}, 104

\bibitem[\protect\citeauthoryear{{NVIDIA Corporation}}{{NVIDIA Corporation}}{2025}]{CUDAGuide}
{NVIDIA Corporation} 2025, {CUDA C++ Programming Guide}, \url {https://docs.nvidia.com/cuda/cuda-c-programming-guide/index.html}

\bibitem[\protect\citeauthoryear{Neal}{Neal}{2003}]{Neal2003_slice}
Neal R.~M.,  2003, \mn@doi [The Annals of Statistics] {10.1214/aos/1056562461}, 31, 705

\bibitem[\protect\citeauthoryear{Owens, Houston, Luebke, Green, Stone  \& Phillips}{Owens et~al.}{2008}]{GPU_computing}
Owens J.~D.,  Houston M.,  Luebke D.,  Green S.,  Stone J.~E.,   Phillips J.~C.,  2008, \mn@doi [Proceedings of the IEEE] {10.1109/JPROC.2008.917757}, 96, 879

\bibitem[\protect\citeauthoryear{Prathaban \& Handley}{Prathaban \& Handley}{2024}]{Prathaban_PE_errors}
Prathaban M.,  Handley W.,  2024, \mn@doi [Monthly Notices of the Royal Astronomical Society] {10.1093/mnras/stae1908}, 533, 1839

\bibitem[\protect\citeauthoryear{Prathaban et~al.}{Prathaban et~al.}{2025b}]{Prathaban_2025_github}
Prathaban M.,  et~al., 2025b, {blackjax\_ns\_gw: blackjax-ns for Gravitational Wave Inference on GPUs}, \url{https://github.com/mrosep/blackjax_ns_gw}

\bibitem[\protect\citeauthoryear{Prathaban, Yallup, Alvey, Yang, Templeton  \& Handley}{Prathaban et~al.}{2025a}]{Prathaban_2025_zenodo}
Prathaban M.,  Yallup D.,  Alvey J.,  Yang M.,  Templeton W.,   Handley W.,  2025a, {Gravitational-wave inference at GPU speed: A bilby-like nested sampling kernel within blackjax-ns}, \mn@doi{10.5281/zenodo.17012011}, \url {https://doi.org/10.5281/zenodo.17012011}

\bibitem[\protect\citeauthoryear{Prathaban, Bevins  \& Handley}{Prathaban et~al.}{2025c}]{Prathaban}
Prathaban M.,  Bevins H.,   Handley W.,  2025c, \mn@doi [Monthly Notices of the Royal Astronomical Society] {10.1093/mnras/staf962}, p. staf962

\bibitem[\protect\citeauthoryear{Pratten et~al.,}{Pratten et~al.}{2021}]{IMRPhenomXPHM}
Pratten G.,  et~al., 2021, \mn@doi [Phys. Rev. D] {10.1103/PhysRevD.103.104056}, 103, 104056

\bibitem[\protect\citeauthoryear{Romero-Shaw et~al.,}{Romero-Shaw et~al.}{2020}]{bilby_validation}
Romero-Shaw I.~M.,  et~al., 2020, \mn@doi [Monthly Notices of the Royal Astronomical Society] {10.1093/mnras/staa2850}, 499, 3295

\bibitem[\protect\citeauthoryear{Skilling}{Skilling}{2006}]{skilling}
Skilling J.,  2006, \mn@doi [Bayesian Analysis] {10.1214/06-BA127}, 1, 833

\bibitem[\protect\citeauthoryear{Smith, Ashton, Vajpeyi  \& Talbot}{Smith et~al.}{2020}]{Smith:2019ucc}
Smith R. J.~E.,  Ashton G.,  Vajpeyi A.,   Talbot C.,  2020, \mn@doi [Mon. Not. Roy. Astron. Soc.] {10.1093/mnras/staa2483}, 498, 4492

\bibitem[\protect\citeauthoryear{Speagle}{Speagle}{2020}]{dynesty}
Speagle J.~S.,  2020, \mn@doi [Monthly Notices of the Royal Astronomical Society] {10.1093/mnras/staa278}, 493, 3132–3158

\bibitem[\protect\citeauthoryear{Storn \& Price}{Storn \& Price}{1997}]{DE}
Storn R.,  Price K.,  1997, \mn@doi [J. Global Optim.] {10.1023/A:1008202821328}, 11, 341

\bibitem[\protect\citeauthoryear{{The LIGO Scientific Collaboration} et~al.}{{The LIGO Scientific Collaboration} et~al.}{2015}]{aLIGO}
{The LIGO Scientific Collaboration} et~al., 2015, \mn@doi [Classical and Quantum Gravity] {10.1088/0264-9381/32/7/074001}, 32, 074001

\bibitem[\protect\citeauthoryear{Thrane \& Talbot}{Thrane \& Talbot}{2019}]{Thrane_2019}
Thrane E.,  Talbot C.,  2019, \mn@doi [Publications of the Astronomical Society of Australia] {10.1017/pasa.2019.2}, 36

\bibitem[\protect\citeauthoryear{Veitch et~al.,}{Veitch et~al.}{2015}]{lal}
Veitch J.,  et~al., 2015, \mn@doi [Physical Review D] {10.1103/physrevd.91.042003}, 91

\bibitem[\protect\citeauthoryear{{Williams}, {Veitch}  \& {Messenger}}{{Williams} et~al.}{2021}]{Williams2021Nessai}
{Williams} M.~J.,  {Veitch} J.,   {Messenger} C.,  2021, \mn@doi [Phys. Rev. D] {10.1103/PhysRevD.103.103006}, \href {https://ui.adsabs.harvard.edu/abs/2021PhRvD.103j3006W} {103, 103006}

\bibitem[\protect\citeauthoryear{Wong, Isi  \& Edwards}{Wong et~al.}{2023a}]{wong2023fastgravitationalwaveparameter}
Wong K. W.~K.,  Isi M.,   Edwards T. D.~P.,  2023a, Fast gravitational wave parameter estimation without compromises (\mn@eprint {arXiv} {2302.05333}), \url {https://arxiv.org/abs/2302.05333}

\bibitem[\protect\citeauthoryear{Wong, Gabri\'e  \& Foreman-Mackey}{Wong et~al.}{2023b}]{flowMC}
Wong K. W.~k.,  Gabri\'e M.,   Foreman-Mackey D.,  2023b, \mn@doi [J. Open Source Softw.] {10.21105/joss.05021}, 8, 5021

\bibitem[\protect\citeauthoryear{Wouters, Pang, Dietrich  \& Van Den~Broeck}{Wouters et~al.}{2024}]{Wouters_BNS}
Wouters T.,  Pang P. T.~H.,  Dietrich T.,   Van Den~Broeck C.,  2024, \mn@doi [Phys. Rev. D] {10.1103/PhysRevD.110.083033}, 110, 083033

\bibitem[\protect\citeauthoryear{Yallup, Handley, Cabezas,   et~al.}{Yallup et~al.}{2025b}]{blackjax_ns_github}
Yallup D.,  Handley W.,  Cabezas J.,    et~al., 2025b, {blackjax}: A Library of Bayesian Inference Algorithms in JAX (Nested Sampling Branch), \url{https://github.com/handley-lab/blackjax/tree/nested_sampling}

\bibitem[\protect\citeauthoryear{Yallup, Kroupa  \& Handley}{Yallup et~al.}{2025a}]{yallup2025nested}
Yallup D.,  Kroupa N.,   Handley W.,  2025a, in Frontiers in Probabilistic Inference: Learning meets Sampling. \url {https://openreview.net/forum?id=ekbkMSuPo4}

\bibitem[\protect\citeauthoryear{Zackay, Dai  \& Venumadhav}{Zackay et~al.}{2018}]{relativebinning2}
Zackay B.,  Dai L.,   Venumadhav T.,  2018, Relative Binning and Fast Likelihood Evaluation for Gravitational Wave Parameter Estimation (\mn@eprint {arXiv} {1806.08792}), \url {https://arxiv.org/abs/1806.08792}

\bibitem[\protect\citeauthoryear{{ter Braak}}{{ter Braak}}{2006}]{DE2}
{ter Braak} C. J.~F.,  2006, \mn@doi [Statistics and Computing] {10.1007/s11222-006-8769-1}, 16, 239

\makeatother
\end{thebibliography}




\appendix

\section{PP coverage plot for \texttt{bilby} and \texttt{dynesty}}
\label{app:bilby_pp_plot}

\begin{figure}
    \centering
    \includegraphics[width=\columnwidth]{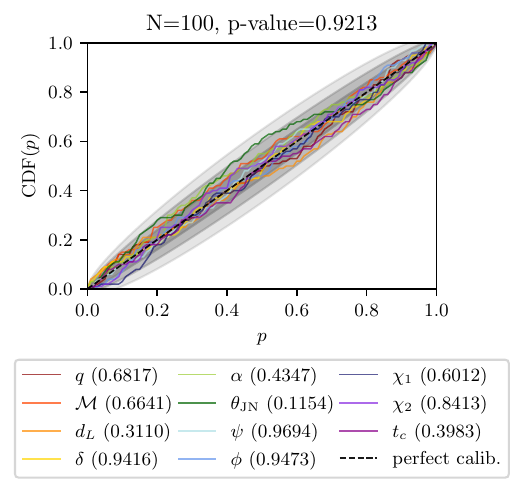}
    \caption{PP coverage plot for the 100-injection study, obtained
    with the CPU-based \texttt{bilby+dynesty} sampler. This plot is
    provided for direct comparison with the results from our
    \texttt{blackjax-ns} implementation shown in
    Figure~\ref{fig:pp_coverage}. The results confirm that the CPU-based
    and GPU-based implementations are functionally similar.}
    \label{fig:pp_coverage_bilby}
\end{figure}

For completeness, we present the PP coverage plot for the CPU-based
\texttt{bilby}+\texttt{dynesty} analysis in
Figure~\ref{fig:pp_coverage_bilby}. This serves as a reference for the
corresponding plot for our \texttt{blackjax-ns} implementation shown in
the main text.

As expected for two functionally equivalent samplers, the resulting PP
plot and p-values demonstrate results consistent with
those from our GPU-based framework. We note, however, that while
the two samplers are algorithmically analogous, apart from differences
in the parallelisation strategy, they will not produce
identical PP plots due to their stochastic nature and the inherent
uncertainties in posterior reconstruction from nested sampling.

The primary sources of this variance include the statistical
uncertainty in the weights assigned to the dead points, which are
themselves estimates~\citep{skilling, Higson_Errors}, and an additional
uncertainty inherent to parameter estimation with nested sampling that
is not captured by standard error propagation
methods~\citep{Prathaban_PE_errors}. Consequently, the PP curves and
their associated p-values are themselves subject to statistical
fluctuations. The results from the two frameworks are therefore only
expected to be consistent within these intrinsic uncertainties.
We do not explicitly account for these uncertainties
here, instead leaving this analysis for future work.

\section{Full run statistics for injection study}
\label{app:injection_study_stats}

\begin{figure*}
    \centering
    \includegraphics{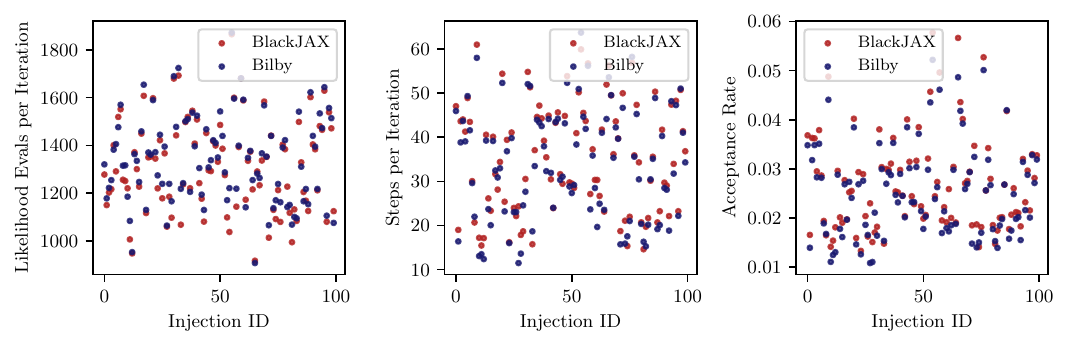}
    \caption{Comparison of internal run statistics for the 100-injection
    study. From left to right: the mean number of likelihood evaluations per iteration,
    the mean number of accepted steps, and the mean acceptance rate, for both the \texttt{bilby} and 
    \texttt{blackjax-ns} runs. The results show that despite the differences in tuning strategy,
    motivated by architectural differences between the CPU and GPU, the two samplers exhibit similar
    properties.}
    \label{fig:injection_study_stats}
\end{figure*}

We present a detailed comparison of the internal sampling statistics for
the 100-injection study in Figure~\ref{fig:injection_study_stats}.
We tuned the `delay' parameter in the walk length tuning formula to
target a similar level of sample decorrelation to \texttt{bilby}.
The results show that this was successful, and both samplers ended up
performing similar numbers of likelihood evaluations, with similar acceptance rates.
Figure~\ref{fig:injection_study_stats} shows the full statistics for each injection.
Although we had to employ batch tuning to the chains, in contrast to the per-iteration
strategy employed in \texttt{bilby}, the results show that this still resulted in 
similar behaviours for the chains and thus our results represent a like-for-like comparison.

\section{Likelihood level parallelisation performance}
\begin{figure}
    \centering
    \includegraphics[width=1.0\columnwidth]{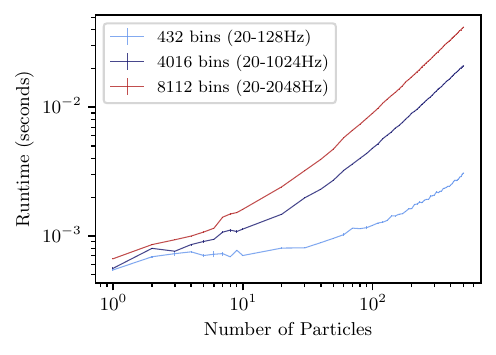}
    \caption{Scaling of likelihood wallclock time on an NVIDIA L4 GPU. 
    A typical representative likelihood for a 3-detector analysis 
    on a 4 second signal using the \texttt{IMRPhenomD} waveform model from 
    \texttt{ripple} is used. Calls of likelihood vectorized over $N$ 
    samples are made, and the wall time averaged over many calls is reported. 
    Comparison is made between different numbers of frequency bins over 
    which the likelihood is evaluated. As can be seen, reducing the number of 
    frequency bins alleviates this bottleneck, and one way to achieve this for
    realistic likelihoods is by using heterodyning.}
    \label{fig:scaling_l4}
\end{figure}

The performance gains from a GPU sampling framework arises from the ability to exploit massive parallelism. 
In particular, we seek to exploit parallelism across multiple short Markov chains, as this is the central 
construction of NS. The potential bottlenecks for parallelism arise from the NS algorithms ability to 
effectively synchronise the chains, and (if this condition is met) the ability to parallelise the likelihood 
evaluation itself across these chains. Developing algorithms that can effectively achieve the former goal 
is part of the ongoing programme of research this work establishes on a GPU GW problem. Regardless of 
effective parallelism across chains, the algorithm in this work is bottlenecked by the ability to 
parallelise the likelihood evaluation itself. We demonstrate the scaling of the likelihood evaluation 
time with the number of samples in Figure~\ref{fig:scaling_l4}. The emerging pattern of polynomial 
runtime scaling with the number of samples, particularly for the high frequency bin number likelihoods we investigate
primarily in this work, indicates that the parallel throughput of the GPU is being saturated. This 
bottleneck can be mitigated by either; using a compressed likelihood (e.g. by heterodyning) or expanding 
processing over multiple GPUs. If possible, rewriting the likelihood to be more parallelism friendly is 
the most efficient solution, but this is challenging for GW data analysis.

\bsp	
\label{lastpage}
\end{document}